\documentclass[aps,prl,reprint,superscriptaddress,twocolumn,nofootinbib,nobibnotes]{revtex4-1}
\pdfoutput=1

\usepackage{amsmath,amssymb,amsthm,mathrsfs,bbm}
\usepackage[dvipsnames]{xcolor}
\usepackage{graphicx}
\usepackage[pdftex, colorlinks=true, citecolor=ForestGreen]{hyperref}
\usepackage{enumitem}

\newcommand{\be}{\begin{equation}}
\newcommand{\ee}{\end{equation}}
\newcommand{\bea}{\begin{eqnarray}}
\newcommand{\eea}{\end{eqnarray}}

\newcommand{\de}{\text{d}}

\begin{document}

\title{\texorpdfstring{Quantum simulation of the Sachdev--Ye--Kitaev model \\ using time-dependent disorder in optical cavities}{Quantum simulation of cSYK using random unitary circuits realised in optical cavities}}

\author{Rahel Baumgartner}
\thanks{These two authors contributed equally.}
\affiliation{Department of Theoretical Physics, University of Geneva, 24 quai Ernest-Ansermet, 1211 Gen\`eve 4, Suisse}
\author{Pietro Pelliconi}
\thanks{These two authors contributed equally.}
\affiliation{Department of Theoretical Physics, University of Geneva, 24 quai Ernest-Ansermet, 1211 Gen\`eve 4, Suisse}
\affiliation{Department of Physics, Princeton University, Princeton, NJ 08544, USA}
\author{Soumik Bandyopadhyay}
\affiliation{Pitaevskii BEC Center, CNR-INO and Dipartimento di Fisica, Università di Trento, I-38123 Trento, Italy}
\affiliation{INFN-TIFPA, Trento Institute for Fundamental Physics and Applications, Trento, Italy}
\author{Francesca Orsi}
\affiliation{Institute of Physics and Center for Quantum Science and Engineering, École Polytechnique Fédérale de Lausanne (EPFL), CH-1015 Lausanne, Switzerland}
\author{Nick Sauerwein}
\affiliation{Institute of Physics and Center for Quantum Science and Engineering, École Polytechnique Fédérale de Lausanne (EPFL), CH-1015 Lausanne, Switzerland}

\author{Philipp Hauke}
\affiliation{Pitaevskii BEC Center, CNR-INO and Dipartimento di Fisica, Università di Trento, I-38123 Trento, Italy}
\affiliation{INFN-TIFPA, Trento Institute for Fundamental Physics and Applications, Trento, Italy}
\author{Jean-Philippe Brantut}
\affiliation{Institute of Physics and Center for Quantum Science and Engineering, École Polytechnique Fédérale de Lausanne (EPFL), CH-1015 Lausanne, Switzerland}
\author{Julian Sonner}
\email{julian.sonner@unige.ch}
\affiliation{Department of Theoretical Physics, University of Geneva, 24 quai Ernest-Ansermet, 1211 Gen\`eve 4, Suisse}

\date{\today}

\begin{abstract}
The Sachdev--Ye--Kitaev (SYK) model is a paradigm for extreme quantum chaos, non-Fermi-liquid behavior, and holographic matter. Yet, the dense random all-to-all interactions that characterize it are an extreme challenge for realistic laboratory realizations. 
Here, we propose a general scheme for densifying the coupling distribution of random disorder Hamiltonians, using a Trotterized cycling through sparse time-dependent disorder realizations. 
To diagnose the convergence of sparse to dense models, we introduce an information-theory inspired diagnostic. 
We illustrate how the scheme can come to bear in the realization of the complex SYK$_4$ model in cQED platforms with available experimental resources, using a single cavity mode together with a fast cycling through independent speckle patterns. 
The simulation scheme applies to the SYK class of models as well as spin glasses, spin liquids, and related disorder models, bringing them into reach of quantum simulation using single-mode cavity-QED setups and other platforms.  
\end{abstract}

\maketitle

\section{Introduction}
Random disorder Hamiltonians form an important entry in the canon of strongly correlated many-body systems, due to their wide applicability and rich phenomenology, ranging from quantum spin glasses \cite{Sachdev:1992fk} over neural networks \cite{hopfield1982neural} all the way to quantum gravity, notably with the advent of the Sachdev--Ye--Kitaev (SYK) class of systems and their nearly conformal holographic states of matter at low temperatures \cite{Sachdev:1992fk,Kitaev-talks:2015,Maldacena:2016hyu}. Nevertheless, concrete proposals for experimental realizations of such models are hard to come by, both in solid-state systems \cite{Pikulin:2017mhj,chew2017approximating,Chen:2018wvb} as well as on quantum-simulation platforms \cite{Danshita:2016xbo,Garcia-Alvarez:2016wem,babbush2019quantum,wei2021optical,Uhrich:2023ddx}. The latter approach has seen some experimental successes, which however, for the time being, remain confined to small system sizes \cite{luo2019quantum,jafferis2022traversable}. On the solid-state side, realizing the platform of irregular graphene flakes as proposed in \cite{Chen:2018wvb}, progress has been reported towards a transition from a Fermi-liquid to an SYK-like phase \cite{PhysRevLett.132.246502}. Virtually all current implementation avenues share the feature that they give rise to `sparse' versions of the desired target models, which are characterised by the property that the disorder is less pronounced than in the idealized target systems. 

\begin{figure}[ht]
\begin{center}
\includegraphics[width=1.0\columnwidth]{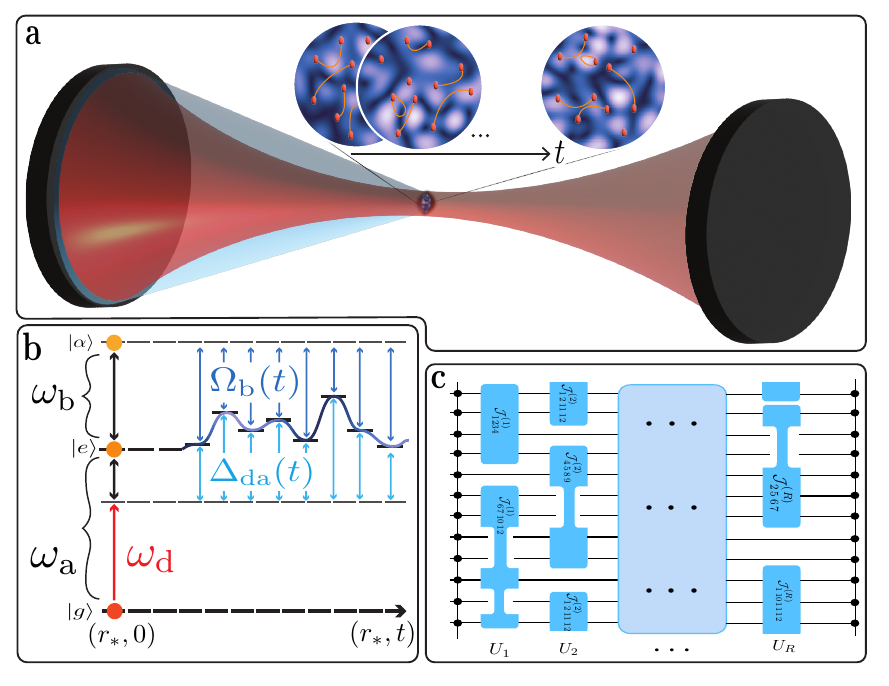}
\caption{
$\textbf{a)}$ Fermionic ${}^6$Li atoms (dark cloud) are trapped at the antinode of a longitudinal mode (red) of a single-mode optical cavity. The system is pumped from the side with a drive (not shown) near-resonant to the cavity field, allowing for atomic scattering from the pump into the cavity and vice versa. Via the use of a light-shift beam (blue), a random phase mask is projected onto the almost two-dimensional trapped cloud of atoms. The resulting speckle pattern with random intensity peaks (white) and valleys (dark blue) is shown above in the zoomed-in circles, together with illustrations of sparse two-body interactions. The speckle pattern is quickly cycled in time through $R$ independent realizations, which enhances the disordered rank-two cavity interactions to SYK-like chaotic interactions. $\textbf{b)}$ Atomic level structure, $|{g;e;\alpha}\rangle$ and respective transition frequencies, $\omega_{\rm a;b}$. Detuning the drive frequency $\omega_{\rm d}$ considerably from $\omega_{\rm a}$ enables adiabatic elimination of the excited states. The speckled intensity $\Omega_{\rm b}$ introduces a position-dependent AC-Stark shift of the excited states and additionally imprints a dependency on time onto the interactions via $\Delta_{\rm da}(t)$. $\textbf{c)}$ 
Random unitary representation of SYK$_4$. Each layer has a randomly selected number of four-Fermi interactions whose couplings are sampled from a sparse distribution. For visual reasons, we do not show couplings that would overlap with one another in a given layer, though in reality these can occur.
\label{fig:cavity}}
\end{center}
\end{figure}

The aim of this paper is two-fold. We describe a novel experimental scheme, based on Trotterizing elementary sparse Hamiltonians, which allows us to densify the systems under consideration in a way that ultimately converges to the fully random target model. This scheme is general and applicable to a large class of models mentioned above. In the process, we also propose a 
quantitative measure of sparseness of a given model with respect to the full model, which may prove to be of use more widely. Building on this idea, we then expand on the main development presented in this paper, namely the realization of a concrete example of a dense disorder Fermionic model. As we show, this model can be implemented in a high-finesse cavity hosting a gas of fermionic lithium atoms interacting with the cavity mode (see Fig.~\ref{fig:cavity}), permitting to produce a simulation of the fully dense (complex) SYK$_4$ model using existing experimental resources. 
The scheme permits to optimally exploit existing experimental resources that can efficiently generate a single layer, which we efficiently combine with a protocol for densifying the interaction, based on rapid switching between individual sparse layers.

\section{c-SYK in c-QED} \label{sec.SYK_in_cavity_QED}

Our main case of interest is the complex SYK$_4$ model of random all-to-all interacting fermions with Hamiltonian 
\begin{equation}\label{eq.FullSYK4model}
    H = \sum_{i<j; k<l}^N J_{ijkl} \, c^\dagger_{i} c^\dagger_{j} c_{k} c_{l} \ ,
\end{equation}
where the couplings are independent random numbers chosen from a Gaussian distribution with vanishing mean value, $\overline{J_{ijkl}} = 0$, and variance $\overline{J^2_{ijkl}} = 2J^2 / N^{3}$. This model has many interesting features, including a maximal scrambling exponent \cite{Kitaev-talks:2015} and an emergent holographic phase at low temperatures \cite{Georges:2001uso}, where its thermodynamic properties assume the same form as those of a near-extremal charged black hole \cite{Sachdev:2015efa} with near-horizon geometry AdS$_2\times \mathbb{R}^{n-2}$. In fact, this extends to a full holographic duality between the theory of nearly-AdS$_2$ gravity and the Schwarzian sector of the SYK model at low temperatures \cite{Maldacena:2016hyu}. 

The model has proven to be a highly fruitful testing ground for ideas in quantum chaos \cite{Rosenhaus:2018dtp,Gu:2019jub} as well as in quantum gravity \cite{Brown:2019hmk}, to cite just two among a large number of fields of interest. 
Consequently, various groups have investigated the possibility of experimental realizations of this class of models 
\cite{Danshita:2016xbo,Garcia-Alvarez:2016wem,Pikulin:2017mhj,chew2017approximating,Chen:2018wvb,luo2019quantum,babbush2019quantum,wei2021optical,jafferis2022traversable,Uhrich:2023ddx,PhysRevLett.132.246502}. 
A commonality between the most promising known approaches is that they typically result in versions of the model(s) that have a reduced amount of randomness, be it due to sparseness of the couplings or the reduced rank of the coupling tensor determining the randomness, ${\cal J}_{ijkl}^{\rm red}$. This has led to a series of interesting investigations to what extent such sparser models still capture the essence of a holographic phase as well as other important properties of the full model \cite{Garcia-Garcia:2020cdo,Kim:2019lwh,Xu:2020shn,Caceres:2023yoj,Orman:2024mpw,tezuka2023binary}. In this work, we choose to take a different route, namely we describe a novel scheme that combines the ideas of random unitary circuits with a Trotterized time evolution, allowing one to use the sparse versions of a given model as basic building blocks to produce the full dense model.

While this approach is widely applicable for the quantum simulation of a range of random disorder models (see Ref.~\cite{longer_paper}), in this paper we focus on a paradigmatic scenario where the effective Hamiltonian is generated as a reduced-rank model with couplings
\begin{equation}\label{eq.reducedRankJtensor}
{\cal J}^{\rm red}_{ijkl} = J_{ik}J_{jl} - J_{jk}J_{il}\,,
\end{equation}
with $J_{ij}$ a random tensor of rank two whose probability distribution is (approximately) Gaussian. 
This situation occurs in several proposals, which effectively restore the full rank by summing over a number $R$ of statistically independent instances of the reduced model, e.g., by addressing simultaneously a large number of cavity modes \cite{Uhrich:2023ddx} or of molecular states \cite{Danshita:2016xbo}, or by decomposing the Coulomb interaction between the lowest Landau levels of a irregular graphene flake, as in \cite{Chen:2018wvb} (see \cite{Kim:2019lwh}). It seems natural that this feature persists for all simulation proposals that engineer SYK terms from a density--density interaction of the constituent fermions. 
The resulting simulated Hamiltonian takes the approximate form 
\begin{equation}\label{eq.SumofReducedHamiltonian}
H_{\rm sim} = \sum_{\alpha=1}^R H_\alpha\,,\quad \textrm{with}\quad H_\alpha = \sum_{i<j; k<l} {\cal J}^{{\rm red}\,,\alpha}_{ijkl}c^\dagger_i c^\dagger_j c_k c_l\,.
\end{equation}
Our aim is, first, to show that in the large $R$ limit $\mathcal J_{ijkl}\equiv \sum_{\alpha} \mathcal J^{\rm red, \alpha}_{ijkl}$ indeed converges to $J_{ijkl}$ of the cSYK model of Eq.~\eqref{eq.FullSYK4model}, up to a factor $R$ in the variance. (To permit a clear comparison, we choose $\overline{({\cal J}^{\rm red}_{ijkl})^2} = 2J^2/N^4$, so that for $R \sim N$ we have $\overline{({\cal J}_{ijkl})^2} = 2J^2/N^3$). 
To this end, we introduce a measure to rigorously quantify this convergence. 
Second, these insights enable us to design a Trotterized scheme where the $H_\alpha$ are realized sequentially rather than simultaneously, thus significantly reducing the experimental overhead. 

Even though our motivation to introduce the model \eqref{eq.SumofReducedHamiltonian} with a lower-rank coupling distribution derives from cavity-QED quantum simulations \cite{Uhrich:2023ddx}, the class of systems described by \eqref{eq.SumofReducedHamiltonian} and its variants are much broader. The central idea applies to any sparse version of the SYK model, the Majorana SYK model\footnote{including SYK$_q$ models in which clusters of $q$ spins interact non-locally, the above example being $q=4$.}, sparsified versions of the Sherrington--Kirkpatrick model \cite{PhysRevLett.35.1792} (see also \cite{Gao:2023gta}) and of the spherical p-spin model \cite{crisanti1992spherical}, combinatorial optimization problems~\cite{Hauke2015}, and a host of other models.

\subsection{Single-mode cQED for low-rank SYK}\label{sec.cavityQED}
As a concrete illustration of the arguments above, we consider the realization of an effective SYK$_4$ model using a single-mode optical cavity hosting a quasi two-dimensional cloud of ultracold trapped ${}^6$Li atoms.  The experimental setup follows the one of the multi-mode proposal of Ref.~\cite{Uhrich:2023ddx}, but with two crucial differences: first, we implement dense disorder without the need for a technically challenging multi-mode cavity and, second, the disorder can---additionally to position dependence---inherit a dependency on time. Employing time-dependent disorder, we implement the discrete time steps of a random unitary circuit using a single-mode cavity, Fig.~\ref{fig:randomCircuit}a. With fast cycling, in the spirit of Trotterization, the circuit dynamics converges to that of the full target model with \textit{static random} disorder, see Fig.~\ref{fig:randomCircuit}b.

For the proposed implementation, we envision driving the atoms off-resonantly to a transition between ground $g$ and an excited state $e$. In addition, a speckle pattern induces a disordered light-shift onto the atoms, and is regularly switched after a short time $\Delta t$. The effect is illustrated in Fig.~\ref{fig:cavity}b: an atom at a fixed position $r_*$ in the cloud will experience a drive--atom detuning $\Delta_{\rm da}(r_*)$ that changes as a function of time. 
To understand the effect on the full many-body system, we focus now on a single time step, during which the detuning will be a disordered function of the spatial position within the cloud, labeled by the two-vector $r$. 
The resulting cavity Hamiltonian derived from \cite{Uhrich:2023ddx} and presented in more detail in \cite{longer_paper} as well as in the Supplementary Material, assumes the form  
\begin{equation}
    H = \Delta a^{\dagger} a + \int \mathrm{d}^2 r \, \frac{\Phi^{\dagger}\Phi}{\Delta_{\rm da}(r)} \psi^{\dagger}(r)\psi(r)\,. 
    \label{cavityHamiltonian}
\end{equation}
Here, $\psi$ represents the ground-state field after adiabatic elimination of the excited state, using $\psi_e(r) =\frac{\Phi}{\Delta_{\rm da}(r)} \psi_g(r)$, where $\Phi=\Omega_{\rm d} + \frac{1}{2}\Omega a $. The single mode of the cavity lives at a detuning $\Delta$ from the drive frequency, is coupled to the atomic state with a single--atom single--photon coupling $\Omega/2$, and is described by annihilation (creation) operator $a$ ($a^\dagger$). At large $\Delta$, the cavity mode can be integrated out, which to second order can be implemented by the Schrieffer--Wolf transformation $e^{-S}H e^S$, generated by $S = - \left( a \Theta/\Delta - h.c. \right)$, where
\begin{equation}
     \Theta = \int \de^2 r \, \frac{\Omega_{\rm d}^* \Omega}{2 \Delta_{\rm da}(r)} \, \psi^\dagger(r) \psi(r) \ .
\end{equation}
The final effective model expanded in the fermionic single-mode basis $ \psi(r) = \sum_i \phi_i(r) \, c_i \ $ and including normal ordering takes the form
\begin{equation}
    H_{\rm eff} =  \sum_{il} M_{il} \, c_{i}^\dagger c_{l} + \sum_{ij, kl} {\cal J}^{\rm red}_{ijkl} \, c_{i}^\dagger c_{j}^\dagger c_{k} c_{l}   \ .
    \label{normal_ordered_hamiltonian}
\end{equation}
The couplings (up to an un-important overall sign) are $M_{il} =  - \sum_{j} J_{ij}  J_{jl}$ and ${\cal J}^{\rm red.}_{ijkl}$ defined as in \eqref{eq.reducedRankJtensor} with
\begin{equation}
    J_{i k} = \frac{\sqrt{\mathcal{E}}}{2} \int \de^2 r \, \frac{\phi_{i}^*(r) \phi_{k}(r)}{\Delta_{\rm da}(r)/\Delta_{\rm da}} \ . 
    \label{Effective_model_couplings}
\end{equation}
Here, we introduced $\mathcal E = |\Omega_{\rm d} \Omega|^2/[|\Delta_{\rm da}|^2 (\omega_c - \omega_{\rm d})]$ as energy scale. The next step in our proposal consists in Trotterizing $H_{\rm eff}$ over different realizations to obtain $H_{\rm sim}$.

\subsection{Trotterized cQED: convergence to SYK}

The phenomenology of the effective model introduced above is rich, behaving identically to cSYK in some aspects but not all. As we argue below, the two models converge for sufficiently large values of $N$, but differ in interesting ways for intermediate values of this parameter. First of all, the integrable mass deformation $\sim M_{il}$ present in \eqref{normal_ordered_hamiltonian} importantly does not contribute significantly at large-$N$, similar to the case of Ref.~\cite{Uhrich:2023ddx}.

Turning to the two-body couplings, we see that the disorder of the model is solely given by the position-dependent detuning $\Delta_{\rm da}(r)$. As a good approximation, we can assume its mean to be position-independent, so that on average it contributes as a constant to the integral. Thus, we can easily see that the couplings $J_{ik}$ at different fermionic modes (called sites henceforth) have vanishing mean, while their mean is non-zero for equal sites, $\overline{J_{ii}}\neq 0$. This also translates into the $\mathcal J^{\rm red.}_{ijij}$ (and permutations), which do not have zero mean. Luckily, all these contribute an overall constant to the Hamiltonian, which we can neglect. On the other hand, for similar reasons, different sets of couplings follow different probability distributions, both in shape and in the variances. Indeed, one can show that a comparatively large constant mean in $\overline{J_{ii}}$ significantly enhances the variance of {\it diagonal couplings} of the form $\mathcal J_{ijil}$, and {\it almost diagonal couplings} of the form $\mathcal J_{ijil}$ (and permutations). By contrast, the only couplings left unaffected are the {\it off-diagonal} ones, namely $\mathcal J_{ijkl}$ with all indices different, which have the smallest variance and are distributed according to a $K_0$ Bessel function. While in the strict large-$N$ limit only the off-diagonal couplings contribute, at intermediate $N$--including values realistically attainable in cQED experiments--the variance in the different coupling sectors affects the dynamics of chaos probes. By explicitly separating the constant mean $\overline{J_{ii}}$, it becomes clear that the issue originates from an additional SYK$_{2}$ contribution that contaminates the chaotic SYK$_4$ dynamics. Fortunately, since this SYK$_{2}$ is linear in the $J_{ij}$ couplings, it can be dynamically compensated by cyclically switching the detuning $\Delta_{\rm da}$ between positive and negative values, thereby eliminating the corresponding term in the Hamiltonian (such technique is commonly called {\it dipole compensation}). 

We have then checked standard chaos probes of scrambling and late-time behavior in the resulting model with the constant $\overline{J_{ii}}$ means removed. In particular we have compared out-of-time-ordered correlators (OTOCs) and the spectral form factor (SFF) in this model with those of SYK$_4$ at different values of $R$, finding that the two converge as $R$ increases. For further details, we refer the reader to \cite{longer_paper} and to Fig.~\ref{fig:randomCircuit}.

\section{Sparse to Dense}\label{sec.sparsetodense}

\subsection{Coupling entropy as a measure of sparseness}
In order to gain a quantitative understanding of the convergence to the target model, in this section we develop measures to characterise the distance between models, and in particular between sparse and dense models. A number of different constructions that give rise to sparsified SYK (or related) models have appeared in the literature, including \cite{Garcia-Garcia:2020cdo,Xu:2020shn,jafferis2022traversable}. In each case, there is a clear procedure and definition of the degree of sparseness of the model, but these measures do not appear to lend themselves to easily compare one type of sparseness with another. In order to remedy this situation, we propose to quantify the degree of sparsification by computing the Shannon information entropy of the relevant coupling distribution, and in particular to determined the relative Shannon entropy (also known as the `KL divergence') with respect to the fully dense model
\begin{equation}
    D\left(P || Q \right) = \int P[X] \log \left( \frac{P[X]}{Q[X]}\right)dX
    \label{eq:relative_Shannon_entropy}\,,
\end{equation}
where $P[X={\cal J}_{ijkl}]$ is the probability distribution of the dense model and $Q[X={\cal J}^{\rm red}_{ijkl}]$ that of the reduced--rank model. If the two distributions are identical, their relative entropy vanishes, so that its approach to zero as a function of a parameter, e.g., the parameter $R$ above, will serve as an important measure allowing to quantify how dense or sparse a given approximation is.

Consider now the case of the low-rank SYK$_4$ model \cite{Kim:2019lwh} introduced in Eq.~\eqref{normal_ordered_hamiltonian} above. For a single disorder realization ($R=1$), when the $J_{ij}$ in Eq.~\eqref{eq.reducedRankJtensor} are Gaussian distributed, the resulting statistical distribution for ${\cal J}^{\rm red}_{ijkl}$ follows a Bessel distribution, 
\begin{equation}
    P_{\rm Bessel}[X] = \frac{1}{\pi \sigma_{(1)} \sigma_{(2)}} \, K_0\left(\frac{|X|}{\sigma_{(1)} \sigma_{(2)} } \right)\ ,
    \label{K_distribution_general}
\end{equation}
where $X={\cal J}^{\rm red}_{ijkl}$ and $\sigma_{(1,2)}^2$ are the variances of the distributions of the rank-two tensors $J_{ij}$. 
We want to tune $\sigma_{(1,2)}$ so that the resulting variance coincides with that of the target model, given by Eq.~\eqref{eq.FullSYK4model}. It may appear complicated to get an analytic expression of the KL divergence of the target Gaussian distribution from a summation over Bessel distributions. However, one can use a characteristic function approach to prove that the latter converges to the former as
\begin{equation}
      D\left(P || Q \right) = \frac{3}{4 R^2} + \mathcal O(R^{-3}) 
\end{equation}
upon summing over $R$. This is an indirect manifestation of the central limit theorem, expressing the convergence rate of $1/R^2$ for their KL divergence.

More generally, one may use similar methods \cite{longer_paper} to show that the probability distribution of the Trotterized sparse model converges to the target distribution with a rate of at least $1/R$, which thus gives a lower bound to the rate of convergence of any observable, including (disorder-averaged) correlation functions.

It is interesting to reconsider a number of approaches to experimentally realise the SYK model and its variants in the literature from this point of view. By inspection of their numerical results, one is often tempted to conclude that the resulting coupling distributions are also given by Bessel distributions (see, e.g., \cite{Chen:2018wvb,Wei:2020ryt,Brzezinska:2022mhj}), hinting at an underlying low-rank or sparsified structure of these proposals. It is certainly heartening that one can envisage to densify these approaches using the approach of this paper.

\subsection{Random unitary circuit SYK}
\begin{figure*}[t]
\begin{center}
\includegraphics[width=1.9\columnwidth]{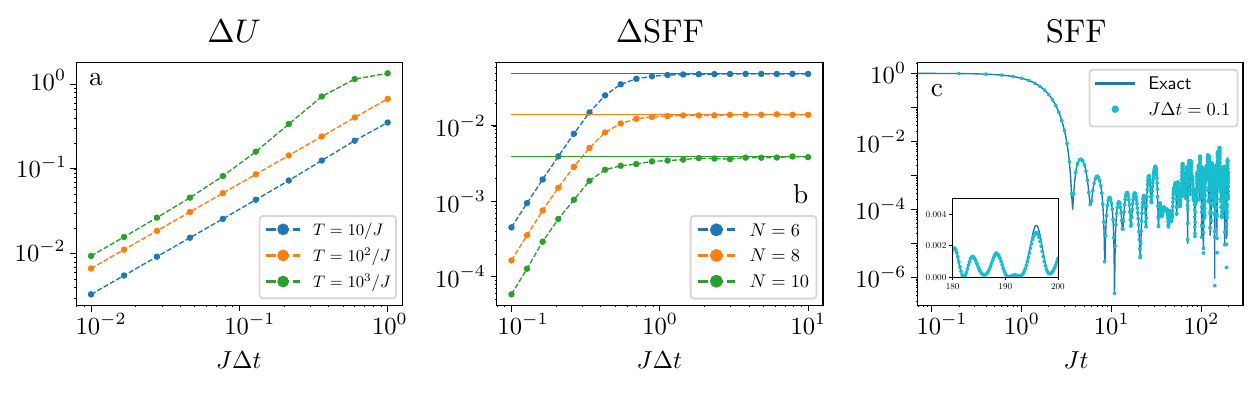}
\caption{ {\bf a)}: Average error \eqref{eq:DU} after Trotterizing the random unitary evolution as a function of $J\Delta t$, for $N = 8$ and different total evolution times. {\bf b)} Average error on the SFF for different values of $N$ and $J \Delta t$, evolving up to $n_{\rm max} = 2000$. {\bf c)} A sample of the SFF for cSYK for $N = 10$ (solid line) and the SFF from the trotterized time evolution \eqref{eq:trotterized_evolution} (bullets). The inset shows a brief period of the evolution at late times.  \label{fig:randomCircuit}}
\end{center}
\end{figure*}
As we discuss below, it is possible to experimentally implement a set of different disorder realizations $H_{\alpha}$ of the sparse version of the model. Here, we  use them as the building blocks of a random unitary circuit, which upon Trotterization converges to the dynamics of the full, i.e., dense, model. Let us start by defining the set of unitaries
\begin{equation}
    U_\alpha = e^{-i H_\alpha \Delta t}
\end{equation}
and $\Delta t$, a sufficiently small time interval. $U_\alpha$ can be thought of as a general quantum gate which acts non-locally on sets of four fermions selected from the total $N$ sites by a single sampling from the random distribution of couplings ${\cal J}^{\rm red}_{ijkl}$. We can then build a random unitary circuit, as shown in Fig.~\ref{fig:cavity}c by successive applications of $R$ different layers, where each layer corresponds to a different sample from $P\left[ {\cal J}_{ijkl} \right]$, labeled ${\cal J}^\alpha_{ijkl}$.

Cycling through the same $R$ disorder patterns $n$ times, a simple application of the BCH formula shows
\begin{equation}
    U_{\rm sim}(t_n) = \left(\prod_{\alpha=1}^R U_{\alpha}(\Delta t)\right)^n + {\cal O}(\varepsilon_n)\,,
    \label{eq:trotterized_evolution}
\end{equation}
where $U_{\rm sim}$ is the time evolution associated to $H_{\rm sim}$, $t_n = n \Delta t$ and the error term is $\varepsilon_n = \frac{t_n \Delta t}{2} \sum_{\alpha < \beta} [H_\alpha,H_\beta] $ up to quadratic order. This implies that in order to simulate the time evolution of a sum of Hamiltonians up to time $t_n$, we can perform $n$ trotterization steps of the $R$ gates $U_{\alpha}$, allowing for a small error. In general, $\Delta t$ has to be smaller than the (inverse) characteristic energy timescale of the system, which in our case is  $1/J$. At the same time, in our specific experimental realization, $\Delta t$ has to be longer than the (inverse) detuning $1/\Delta$, such that the adiabatic elimination can be performed at each Trotter step. To have a quantitative understanding of the error $\varepsilon_n$, we need to estimate the expected value of the commutators. For the case of a low-rank SYK$_4$ model \cite{Kim:2019lwh}, a suitable measure of the size of the error terms is given by the Frobenius operator norm of the sum of commutators, which one can estimate as $\overline{ \|\sum_{\alpha < \beta} [H_{\alpha}, H_{\beta}] \|^2} \lesssim 2 \times 10^{2} J^4 R^2/N^2 + \mathcal O(R^2/N^3)$, where the overline indicates that a disorder average has been performed \cite{longer_paper}. A more precise understanding can be obtained, by computing the average norm of the distance of the Trotterized random unitary evolution operator from the target, i.e.,
\begin{equation}
    \Delta U = \frac{1}{n_{\rm max}} \sum_{n = 1}^{n_{\rm max}} \overline{\Bigg \| U_{\rm sim}(t_n) - \Bigg( \prod_{\alpha = 1}^{R} e^{-i H_\alpha \Delta t} \Bigg)^n \Bigg\|} \ ,
    \label{eq:DU}
\end{equation}
with the total time of the simulation being $T= n_{\rm max} \Delta t$. The results, shown in Fig.~\ref{fig:randomCircuit}, show a convergence for arbitrary time evolutions $T$, provided a sufficiently short $\Delta t$. Interestingly, this conclusion holds also for values of $T$ such that $T \Delta t J^2$ is not small, which are not captured by the $\varepsilon_n$ mentioned above. This can be also checked at the level of observables depending on the time evolution, such as OTOCs and the SFF. See panel b and c of Fig.~\ref{fig:randomCircuit} and \cite{longer_paper}.

\section{Dissipation and Lindbladian SYK}
We now address dissipation in the cavity simulation (see also Supplementary Material). This is an important issue, since it sets the relevant timescales after which the closed-system chaotic dynamics is no longer a good description. The main sources of dissipation in the cavity are photon-scattering and photon-loss. We employ a Markovian approximation, and the corresponding jump operators can be found in \eqref{Lindblad1}--\eqref{Lindblad2} and in \cite{Uhrich:2023ddx}. Schematically, they are of the form $L = \sum_{ij} K_{ij} c_i^\dagger c_j$, coming from the three-body QED interaction vertex between the photon field and the cSYK sites. The coefficients $K_{ij}$ are random, but depend on the speckle pattern used to generate the random couplings $J_{ijkl}$. One would thus expect correlations between the two. In \cite{longer_paper}, we show that in the large-$R$ limit, the two probability distributions decorrelate, so that $P[J_{ijkl}, K_{mn}] = P_{J}[J_{ijkl}] \times P_{K}[K_{mn}] + \mathcal O(1/R)$. This feature allows us to drastically simplify the problem, treating the closed-system dynamics and dissipation independently. This case was analyzed in \cite{Ryu_2022}, where the authors studied how these random Lindblad operators affect the chaotic dynamics in the large-$N$ limit. In particular, employing the Schwinger--Keldysh effective action, it is possible to study the effect of dissipation in the Schwinger--Dyson equations for the two-point function $G(t) = \langle c_i^\dagger(t) c_i(0) \rangle$, and the self energy $\Sigma(t)$. In the Supplementary Material, we show that the effect of dissipation enters in the calculation of $\Sigma(t)$ at an order $\mathcal O(R / N)$, the same large-$N$ order as the unitary evolution. This implies that, even in the large-$N$ limit, we can expect the relative importance between unitary and dissipative dynamics to be controlled by the cooperativity. Below, we give some estimates about experimentally feasible timescales.

\section{Conclusions}
In this paper, we have described a hybrid digital--analog approach to quantum simulate a large class of quenched disorder models, including as the main system of interest a thorough study of the complex SYK$_4$ model, of relevance to the phenomenology of quantum chaos as well as a potential candidate to exhibit a holographic state of matter. We have demonstrated that one can treat cQED-native but notably sparse (or low-rank) realizations of quenched disorder models as generalized quantum gates, which can be Trotterized into the dense target model in a controlled way. We have introduced a measure in the form of the KL divergence to rigorously quantify the approach to the target model. In the specific case of low-rank SYK models, we have shown how the dense disorder of the full-rank model is reached Trotterizing low-rank Hamiltonians, at the level of couplings, time evolution and dynamical observable.

Furthermore, this method of simulating SYK appears to be experimentally feasible in current cQED platforms. 
Control over the light-matter coupling in space and time, close to that introduced in \eqref{cavityHamiltonian}, has been demonstrated recently \cite{orsi:2024aa}. The Trotter-step duration available there is determined by the speed at which independent speckle patterns can be alternated in time. Currently available methods based, e.g., on acousto-optical light modulation \cite{liu:2023ab} yield timescales of the order of $\mu$s, much slower than $1/\Delta$ but fast enough to allow for thousands of steps within an experimental cycle time. The number of independent patterns that can be created optically is given by the number of independent optical modes available over the atomic cloud. Requiring the number of modes to exceed the number of particles thus amounts to having an optical resolution below the Fermi wavelength. This would be achieved in the cavity-microscope system \cite{orsi:2024aa} for a typical mesoscopic Fermi gas, such as reported in \cite{holten:2022aa}. As discussed above, the main limitation will arise from the finite cooperativity, specifying the rate of photon leakage with respect to coherent, virtual photon exchanges \cite{Tanji-Suzuki:2011ac}. The ultimate limit set by fabrication limitations for optical mirrors suggests that cooperativities of order of a few hundreds could be reached with currently available technologies. Cooperativities on that scale have already been reported at optical wavelengths in microcavities \cite{baghdad:2023aa}. There, we expect the SYK regime where $N \sim R$ to be reachable using Fermi gases of up to about $50$ atoms, with timescales reaching the scrambling time $\sim \log(N)/J$.

The methods presented in this work are also interesting if analyzed under the lens of computational complexity. Importantly, the most onerous contribution to the complexity of simulation is removed by engineering the cavity to produce the sparse model intrinsically. The complexity of simulation is thus dominated by the number of Trotter steps required for a given accuracy $\epsilon$. To reach a desired accuracy for a given simulation time $T$, one can show \cite{longer_paper} that a number of steps $n_{\rm Trotter} > M T^2 J^2 R/ ( N \epsilon)$ are required, for $M$ an order one number, which compares very favorably with previous digital and hybrid approaches  \cite{Garcia-Alvarez:2016wem,Xu:2020shn}.

\section{Acknowledgements}
We wish to thank Rohit Prasad Bhatt, Michael Eichenberger, Ekaterina Fedotova, Andrea Legramandi, Pranjal Nayak, David Pascual Solis, Adrián Sánchez-Garrido, Alex Windey, and Yi-Neng Zhou for helpful conversations and comments. This work is supported by the Fonds National Suisse de la Recherche Scientifique through Project Grant $200021\_{}215300$ and the NCCR ``The Mathematics of Physics (SwissMAP)" (NCCR51NF40-141869.)
This work was supported by the Swiss State Secretariat for Education, Research and Innovation (SERI) under contract number UeMO19-5.1 and MB22.00063. 
This project has received funding from the Italian Ministry of University and Research (MUR) through the FARE grant for the project DAVNE (Grant R20PEX7Y3A). 
Funded by the European Union - NextGenerationEU, Mission 4 Component 2 - CUP E53D23002240006 and CUP E63C22000970007. 
This work received funds from project DYNAMITE QUANTERA2\_00056 funded by the Ministry of University and Research through the ERANET COFUND QuantERA II – 2021 call and co-funded by the European Union (H2020, GA No 101017733). 
This project has received funding from the European Union’s Horizon Europe research and innovation programme under grant agreement No 101080086 NeQST.
Funded by the European Union. Views and opinions expressed are however those of the author(s) only and do not necessarily reflect those of the European Union or the European Commission. Neither the European Union nor the granting authority can be held responsible for them.
This work was supported by the Provincia Autonoma di Trento, and Q@TN, the joint lab between University of Trento, FBK—Fondazione Bruno Kessler, INFN—National Institute for Nuclear Physics, and CNR—National Research Council.

\bibliographystyle{utphys}
\bibliography{extendedrefs}{}

\clearpage
\onecolumngrid

\setcounter{page}{1}
\setcounter{equation}{0}
\setcounter{figure}{0}
\setcounter{section}{0}
\renewcommand{\thepage}{S\arabic{page}}
\renewcommand{\theequation}{S\arabic{equation}}
\renewcommand{\thefigure}{S\arabic{figure}}
\renewcommand{\thesection}{S\Roman{section}}
\renewcommand{\figurename}{SUPPLEMENTARY FIG.}

\section{Supplementary Material}

In this Section, we expand on the experimental realization of the effective cavity Hamiltonian, and on dissipation effects. In particular, we give a complete explanation of the various terms that compose the cavity Hamiltonian, and show how they can effectively describe a cSYK$_{4}$ upon integrating out the cavity photons. Moreover, we analyze the various sources of dissipation that affect the cavity implementation, and show how they compete with the unitary evolution computing the Schwinger--Dyson equations of the system.

\subsection{Effective Cavity Hamiltonian}\label{supp:cavity}
In this section, we give more information on how to derive the cavity Hamiltonian presented in \eqref{cavityHamiltonian}. The microscopic many-body Hamiltonian realised in the envisioned cavity setup is composed of many terms, mentioned above and in \cite{Uhrich:2023ddx}. In particular, we have 
\begin{equation}
    H = H_{\rm kt} + H_{\rm c} + H_{\rm a} + H_{\rm ac} + H_{\rm ad} \ .
\end{equation}
The first contribution, $H_{\rm {kt}}$, describes the collective dynamics of the atoms in the center-of-mass frame. These are represented by the field operators $\psi_s(r)$, where $s \in \{ g, e \}$, describing the ground and excited state, respectively. With this notation, we have
\begin{equation}
    H_{\rm kt} = \sum_{s \in \{ e,g\}} \int \de^2r \, \psi_s^{\dagger}(r) \left( \frac{p^2}{2 m_{\rm at}} + V_t(r) \right) \psi_s(r) \ .
\end{equation}
The integral is two dimensional since we are neglecting any movement on the direction of the cavity axis and runs over the plane of the atomic cloud. In the formula above, $m_{\rm at}$ is the atomic mass (we will use specific values for ${}^6$Li atoms), and $V_{t}(r)$ represents the harmonic trapping potential. The second term, $H_{\rm c}$, is the contribution of transverse cavity modes (TCM), written in the rotating frame (see, e.g., \cite{Mivehvar_2021} for a recent review of cQED with quantum gases) of the drive $\omega_{\rm d}$, 
\begin{equation}
    H_{\rm c} = \sum_m (\omega_m -\omega_{\rm d}) a_m^\dagger a_m .
\end{equation} 
The frequency of the different TCM's is $\omega_m = \omega_0 + m \, \delta\omega$, and in the limit $\delta \omega / \omega_0 \gg 1$, we can neglect all higher modes with $m>0$. This regime is within the capabilities of existing laboratory setups \cite{Brantut_rand_spin_model_2023}. The operators $a_m, a_m^{\dagger}$ annihilate respectively create a cavity photon with mode frequency $\omega_m$. For further convenience, we also define $\Delta \equiv \omega_0 -\omega_{\rm d}$. Next, the term $H_{\rm a}$ represents the contribution of the excited state energy of the atoms, 
\begin{equation}
    H_{\rm a} =  \int \de^2 r \, \big( \omega_{\rm a} - \omega_{\rm d} \big) \, \psi_e^{\dagger}(r)\psi_e(r) \ .
\end{equation}
The frequency $\omega_{\rm a}$ corresponds to the transition frequency from ground $|g\rangle$ to excited state $|e\rangle$. We define the detuning $\Delta_{\rm da} \equiv \omega_{\rm a} - \omega_{\rm d} $, and we assume it to be the dominant energy-scale, in order to adiabatically eliminate excited states, leaving the ground state of the $N$ atoms to represent $N$ spinless fermions of the target model, $c_i^\dagger, c_i$. This detuning is actually position dependent, $\Delta_{\rm da}(r)$, due to the position-dependent AC Stark shift induced by the speckle pattern. Finally, the last two terms (written in the rotating frame approximation) are
\begin{equation}
    H_{\rm ac} =  \frac{\Omega}{2} \int \de^2 r \, \left( g_0(r) \psi_e^{\dagger}(r)\psi_g(r) \, a + h.c. \right) \ , \qquad \qquad H_{\rm ad} = \Omega_{\rm d} \int \de^2 r \, \left(  g_{\rm d}(r) \psi_e^{\dagger}(r)\psi_g(r)  + h.c. \right) \ ,
    \label{H_ac_and_H_ad}
\end{equation}
and represent the interactions between the atoms and cavity, respectively between the atoms and drive modes. These are the QED interaction vertices between charged particles and light, suitably divided between the interaction with the classical field (the drive, from $H_{\rm ad}$), and the quantum field, represented by photons in the lowest mode. $\Omega$ and $\Omega_{\rm d}$ are the coupling strengths between the atoms and the cavity mode, and the atoms and the drive, respectively. Further, $g_0(r)$ and $g_d(r)$ are the amplitudes of the cavity and drive modes, which in the two-dimensional plane of the atomic cloud are given by Hermite--Gauss modes. In the main text, we set $g_{\rm d}(r)$ and $g_0(r)$ to unity, the former in a long-wavelength approximation and the latter so that all $J_{ii}$ couplings have (approximately) the same mean. More details can be found in \cite{longer_paper}.

The position-dependent adiabatic elimination of excited states is used to induce disorder in the system, as can be seen in \eqref{cavityHamiltonian}. In order to imprint it, we consider a third auxiliary state $|\alpha \rangle$, which is a state further excited from $|e \rangle$ by the introduction of the light-shift beam of Rabi frequency $\Omega_{\rm b}$. As such, this light-shift beam causes an AC-Stark shift of the energy of the excited state $|e\rangle$, which in turn induces a shift of the detuning frequency $\Delta_{\rm da}$. Allowing this light-shift to be modulated spatially results in a situation where the effective detuning is now a spatially varying function $\Delta_{\rm da}({r})$, resulting in a disordered structure that filters down via the expansion $ \psi(r) = \sum_i \phi_i(r) \, c_i \ $ to the two-body interaction. This results in the effective Hamiltonian \eqref{normal_ordered_hamiltonian}. For more detail, see also the multi-mode proposal \cite{Uhrich:2023ddx}, as well as the companion paper \cite{longer_paper}.

\subsection{Dissipation}

The two main sources of non-Hermitian dynamics in the cavity are two, namely physical photons leaving the cavity, and spontaneous decay from the excited states. These are encoded in the two Hamiltonians of \eqref{H_ac_and_H_ad}, and they represent corrections to integrating out the photons, and to the adiabatic elimination of the excited states. Generally, in a Markovian approximation, the Lindblad jump operator entering the Liouville dynamics is given by the interaction vertex between the system and the environment. In our case, these are \cite{Uhrich:2023ddx}
\begin{align}
    L_{\Gamma} (r) = & \, \frac{\sqrt{\Gamma} \, \Omega_{\rm d}}{\Delta_{\rm da}(r) + i \Gamma/2} \,  g_{\rm d} (r) \psi_{\rm g}^{\dagger}(r) \psi_{\rm g} (r) \label{Lindblad1} \\
    L_{\kappa} (r) = & \, \frac{\sqrt{\kappa} \, \Omega_{\rm d} \Omega}{2(\Delta - i \kappa/2)} \int \de^2 r \,  \frac{g_{\rm d} (r) g_{0} (r)}{\Delta_{\rm da}(r) + i \Gamma/2} \, \psi_{\rm g}^{\dagger}(r)\psi_{\rm g} (r) \label{Lindblad2}
\end{align}
the former modeling photon scattering and the latter photon losses. We then notice that the Lindblad jump operators are quadratic in the fermionic sites, and they depend on the speckle pattern that also generates the $J_{ij}$'s. Therefore, in our case, the jump operators can schematically be written as
\begin{equation}
    L = \sum_{ij} K_{ij} c_i^\dagger c_j \ , \qquad \qquad \overline{|K_{ij}|^2} = \frac{K^2}{N^2} \ ,
\end{equation}
where the $N$-scaling is set to be equal to the $J_{ij}$, for the aforementioned reason. For a more precise explanation we refer the reader to \cite{Uhrich:2023ddx, longer_paper}. 
In both cases, the Lindblad operators arise because of light--matter interaction, and therefore depend on the speckle pattern so that $K_{ij}$ is correlated with $ J_{ij}$, and thus such Lindblad couplings are correlated with the cSYK couplings $\mathcal J^{\rm red.}_{ijkl}$. However, as we mentioned in the main text, in the large $R$ limit, they effectively become uncorrelated up to $1/R$ corrections. 

Neglecting for simplicity this correction, we can take inspiration from \cite{Ryu_2022} to study how this non-Hermitian dynamics affects the system. For simplicity, we restrict to infinite temperature, which is a fixed-point of the Lindblad evolution. Then, in terms of coherent state path integrals, the Lindblad evolution of the density matrix can be rewritten in terms of a Keldysh effective action \cite{Sieberer_2016} as\footnote{See also \href{https://boulderschool.yale.edu/sites/default/files/files/20210712_Boulder-reduced.pdf}{this} set of lecture notes.}
\begin{multline}
    Z  = \int \mathcal D[\psi_\pm, \psi_\pm^\dagger] e^{iS[\psi_\pm, \psi_\pm^\dagger]} \ , \qquad \qquad  S[\psi_\pm, \psi_\pm^\dagger] = \int \de t \bigg[ \psi_+^\dagger i \partial_t \psi_+ - \psi_-^\dagger i \partial_t \psi_- - H_+[\psi_+, \psi_+^\dagger] + H_-[\psi_-, \psi_-^\dagger] \\
    - i \sum_\alpha \left( 2 L_{\alpha, -}^\dagger L_{\alpha,+} - L_{\alpha,+}^\dagger L_{\alpha,+} - L_{\alpha, -}^\dagger L_{\alpha,-} \right) \bigg]  \ .
\end{multline}
The effective action above is quadratic in the jump operators. In order to linearize it, we can employ complex Hubbard--Stratonovich auxiliary fields $\theta_{\alpha, \pm}$ and $\bar \theta_{\alpha, \pm}$. The result is
\begin{multline}
    S[\psi_\pm, \psi_\pm^\dagger, \theta_{\alpha, \pm}, \bar \theta_{\alpha, \pm}] = \\
    \int \de t \left( \psi_+^\dagger i \partial_t \psi_+ - \psi_-^\dagger i \partial_t \psi_- - H_+[\psi_+, \psi_+^\dagger] + H_-[\psi_-, \psi_-^\dagger] \right) +i \int \de t \sum_{\alpha} \left[ \begin{matrix}
    \bar \theta_{\alpha, +} (t) \\  \bar \theta_{\alpha, -} (t)
    \end{matrix} \right]^\mathsf{T} \left[ \begin{matrix}
    1 & 0 \\
    2 & 1
    \end{matrix} \right]
    \left[\begin{matrix}
    \theta_{\alpha, +} (t) \\
    \theta_{\alpha, -} (t)
    \end{matrix} \right] \\
    -i  \int \de t \sum_{\alpha} \Big( \bar \theta_{\alpha, +} (t) L_{\alpha, +} (t) +  \bar \theta_{\alpha, -} (t) L_{\alpha, -} (t) +  L_{\alpha, +}^{\dagger} (t) \theta_{\alpha, +} (t) + L_{\alpha, -}^{\dagger} (t) \theta_{\alpha, -}(t) \Big)  \ .
    \label{SK_action}
\end{multline}
It is not hard to verify that integrating out the auxiliary fields we obtain the original Lindblad evolution. We can now simply derive the Schwinger--Dyson equations for the two-point function $G(t) = \langle c_i^\dagger(t) c_i(0) \rangle$ and the self-energy $\Sigma(t)$, as well as for the corresponding ones of the auxiliary fields $G_{ab}^\theta(t) = \langle \bar \theta_{\alpha, a} (t)  \theta_{\alpha, b} (0) \rangle$ and $\Sigma^\theta_{ab}(t)$. The Schwinger-Dyson equations for the auxiliary fields are
\begin{equation}
    \Sigma_{ab}^{\theta}(t) = - \frac{K^2}{4} G_{ab}^2(t) \ , \qquad \qquad \mathbf{G}^{\theta}(t) = \left[ \begin{pmatrix}
    1 & 0 \\
    -2 & 1\\
    \end{pmatrix} \delta(t) - \mathbf \Sigma^{\theta}(t)\right]^{-1} \ ,
\end{equation}
while the ones for the fermionic fields are
\begin{equation}
    \Sigma_{ab}(t) = \, - \frac{J^2 R}{N} \, s_{ab} G_{ab}^3 (t) + \frac{K^2 R}{N} \left(G_{ab}^\theta (t) + G_{ba}^\theta (-t) \right) G_{ab} (t) \ , \qquad \qquad \mathbf{G}(t) = \left[ \mathbf G_0^{-1}(t) - \mathbf \Sigma(t)\right]^{-1} \ .\label{Sigma_dissipation} 
\end{equation}
The subscripts $a$ and $b$ take values in the set $\{ +, - \}$ with the matrix $s_{ab}$ having components $s_{++} = s_{--} = 1$ and $s_{+-} = s_{-+} = -1$. The inverses refer to both the $\{a,b\}$ indices, as well as for the time-domain.

Equation \eqref{Sigma_dissipation} makes it apparent that dissipation contributes at the same order as the unitary evolution in the large-$N$ limit. Given that chaotic timescales are parametrically large\footnote{for instance, the scrambling time is $t_{\rm scr} \sim \log(N)$.}, this implies that in the large-$N$ limit such chaotic dynamics eventually have to compete with the dissipation scale, and their experimental observability depends on the quality factor of the cavity at a given atom number.

\end{document}